\documentstyle[twoside,fleqn,espcrc2]{article}
\begin{document}
%
\hyphenation{author another created financial paper re-commend-ed}
\title{A new algorithm for numerical simulation of Langevin equations }  
\author{H.Nakajima 
\address{Department of Information Science, Utsunomiya University, Utsunomiya 321, Japan  \\
E-mail address: nakajima@infor.utsunomiya-u.ac.jp}                           
   and
S.Furui \address{School of Science and Engineering, Teikyo University, Utsunomiya 320, Japan\\
E-mail address: furui@dream.ics.teikyo-u.ac.jp}}

\begin{abstract}
Formulated is a new systematic method for obtaining higher order
corrections in numerical simulation of stochastic differential equations
(SDEs), i.e., Langevin equations. Random walk step algorithms within a given
order of finite $\Delta t$, are obtained so as to reproduce within that order
a corresponding transition density of the Fokker-Planck
equations, in the weak Taylor approximation scheme \cite{KP}.
A great advantage of our method is its
straightforwardness such that direct perturbative calculations produce
the algorithm as an end result, so that the
procedure is tractable by computer. Examples in general form for curved
space cases as well as flat space cases are given in some order of
approximations. Simulations are performed for specific examples of U(1) system
and SU(2) systems, respectively.
\end{abstract}

\maketitle

\section{Introduction}

We propose a new systematic algorithm for obtaining higher order
corrections in numerical simulation of SDEs, i.e., Langevin equations
in either flat or curved space.

The problem is how to realize random walk step algorithms for
$\Delta t$, so as to reproduce proper corrections within a given order
$O(\Delta t^n)$ of corresponding transition density which evolves according to
the Fokker-Planck equation.

Our systematic algorithm consists of all direct perturbative calculations
within some order $O(\Delta t^n)$ such as normal
ordering of the time evolution operator, exponentiation of polynomials,
completing squares of polynomials, and inversion of polynomials. These direct
perturbative calculations produce the random walk step algorithm as an end
result without any matching equations, so that the procedure is tractable by
computer.

It is to be noted that Drummond et al.\cite{DDH} assumed linear step algorithms
with respect to random variates, but we do not. In their scheme, equations
becomes overconstrained and will have no solutuion in higher orders,
since otherwise the obtained algorithm generates Gaussian distribution for the
transition density, which is, however, generally impossible in higher orders.

\section{Formalism}

The Langevin equation in d dimensional curved space with
an initial condition $x=x_0$ at $t=t_0$, reads as
\begin{equation}
\partial_t x^i = u^i(x) + e^i_\alpha(x) \eta^\alpha (t)
\qquad(i,\alpha=1,2,\cdots,d),  \end{equation}
\begin{equation}
<\eta^\alpha(t) \eta^\beta(t')>=2\sigma^{\alpha \beta}\delta(t-t'),
\end{equation}
and
\begin{equation}
<\eta^\alpha(t)>=0,
\end{equation}
where (1) is understood as Ito's SDE.
Then we can derive the
corresponding Fokker-Planck equation for the transition density in the
curved space with an initial condition $P(t_0,x;t_0,x_0)=\delta(x-x_0)$ as
\begin{equation}
\partial_t P(t,x;t_0,x_0) =  K P(t,x;t_0,x_0) ,
\end{equation}
where
\begin{equation}
K = \partial_i(\partial_j g^{ij}(x) - u^i(x))   ,       
\end{equation}
and
\begin{equation}
g^{ij} = {e^i}_\alpha {e^j}_\beta \sigma^{\alpha \beta} .
\end{equation}

Covariant form of the Fokker-Planck equation can be derived by
identification, $\phi=P/ \sqrt {det g_{ij}} $ and in case that
\begin{equation}
u^i(x)=(\partial_jg^{ij}(x)P_0(x))/P_0(x) ,
\end{equation}
$P(x)$ tends to a unique asymptotic distribution $P_0(x)$.

Given a $u^i(x)$, we are to find a random walk step algorithm $\Delta x^i$, a function of random variates $\eta^\alpha$, so as to satisfy within a given order $O(\Delta t^n)$,
\begin{equation}
P(t_0+\Delta t,x;t_0,x_0)=<\delta(x-(x_0+\Delta x))>    
\end{equation}
where 
\begin{equation}
<\eta^\alpha,\eta^\beta> = 2\delta^{\alpha \beta}\Delta t,
\end{equation}
and 
\begin{equation}
<\eta^\alpha>=0. 
\end{equation}

\subsection{A new systematic algorithm for higher order random walk step
algorithms}

First we write a truncated exponential operator of the $O(\Delta t^n)$
formal solution
\begin{equation}
e^{\Delta tK}\delta(x-x_0)
\end{equation} 
in normal order, i.e., differential operators in the leftmost
position, and then all velocity and metric fields and
their derivatives may be evaluated at the position $x_0$ and become commutable
with differential operators. Then we insert
\begin{equation}
e^{-\Delta t g^{ij}(x_0)\partial_i\partial_j} e^{\Delta t g^{ij}(x_0)
\partial_i\partial_j }
\end{equation}
between $e^{\Delta tK}$ and $\delta$, and make use of
\begin{equation}
  e^{\Delta t g^{ij}(x_0)\partial_i\partial_j}\delta (x-x_0)={1\over 
\sqrt{(4\pi\Delta t)^d g_0}} e^{-{1\over 4} g_{ij}(x_0)z^i z^j} ,
\end{equation}
where $z^i=(x^i-x_0^i)/\sqrt{\Delta t}, g_0=det(g^{ij}(x_0))$, and 
$\partial_i=\partial_z^i/\sqrt{\Delta t}$ to obtain the transition density
in a form as
\begin{eqnarray}
 &&P(t_0+\Delta t,x;t_0,x_0)=F(\sqrt{\Delta t},z,\cdots)\nonumber\\
&&\times{1\over \sqrt{(4\pi\Delta t)^d g_0}}e^{-{1\over 4} g_{ij}(x_0)z^i z^j}.
\end{eqnarray}

We use a singular coordinate $z^i=(x^i-x_0^i)/\sqrt {\Delta t}$ due to a
singular nature of the perturbation expansion.

Within the given order, we exponentiate the polynomial factor
$F(\sqrt{\Delta t},z,\cdots)$, and proceed perturbatively order by order so as
to eliminate the leading order corrections in the exponent polynomial by
finding proper transformation, that is, by completing squares, to obtain
\begin{equation}
  P(t_0+\Delta t,x;t_0,x_0)={N\over \sqrt{(4\pi \Delta t)^d g_0}}
 e^{-{1\over 4}g_{ij}(x_0)\tilde z^i \tilde z^j},
\end{equation}
where $\tilde z^i$ is a polynomial of $z^j, \tilde z^i=\tilde z^i(\sqrt t,z)$
such that $\tilde z^i=z^i + c^i +\cdots$, and $N$ is a normalization constant
independent of $z$.
Using the $\delta-$function, we can write (15) as
\begin{eqnarray}
 &&P(t_0+\Delta t,x;t_0,x_0)={N\over \sqrt{(4\pi\Delta t)^d}}\nonumber\\
 &&\times \int d\tilde \eta e^{-{1\over 4}\tilde\eta^2} \delta^d(\tilde z^i-\theta^i) ,\end{eqnarray}
where $\theta^i=\sqrt{g^{ij}(x_0)}\tilde\eta^j$, and inverting the argument of the $\delta-$function
with respect to $z^i$, we rewrite the $\delta-$function as 
\begin{equation}
\delta^d (\tilde z^i-\theta^i)=det(dz/d\theta)\delta^d (z^i - z^i(\theta))
\end{equation}
to obtain
\begin{eqnarray}
 &&P(t_0+\Delta t,x;t_0,x_0)={N\over \sqrt{(4\pi)^d}} \int d\tilde \eta 
det(dz/d\theta)\nonumber \\
&&\times e^{-{1\over 4}\tilde\eta^2} \delta^d (x^i-x_0^i-\Delta x^i(\theta)) ,  
\end{eqnarray}
where $\Delta x^i=\sqrt{\Delta t}z^i(\theta)$. The final part of our procedure
to reach the goal in the form
\begin{eqnarray}
&&P(t_0+\Delta t,x;t_0,x_0)={1\over \sqrt{(4\pi)^d}} \int d\eta e^{-{1\over 4}\eta^2}\nonumber\\ &&\times \delta^d (x^i-x_0^i-\Delta x^i(\theta)) ,
\end{eqnarray}
is to find a transformation $\tilde\eta^i=\tilde\eta^i(\eta)$ such that 
\begin{equation}
{N\over \sqrt{(4\pi)^d}} d\tilde \eta det(dz/d\theta) e^{-{1\over 4}\tilde\eta^2}={1\over \sqrt{(4\pi)^d}} d\eta e^{-{1\over 4}\eta^2} 
\end{equation}
which we prove to be always possible as follows.

Algorithm to find the transformation from $\tilde\eta^i$ to $\eta^i$:
 Noting that
\begin{equation}
det(dz/d\theta)=1+\cdots,
\end{equation}
we exponentiate the determinant factor, and consider its exponent in a
form of polynomial of $\tilde\eta$  as $-1/4 (\tilde\eta^2 + \cdots)$. 
We assume that constant terms
are absent since they can be factored out as normalization constant, and also
that either linear or quadratic terms with respect to $\tilde \eta$ in the 
lowest leading
order correction terms are absent since they can be eliminated by some linear
transformation which only effects a constant Jacobian factor. Let us suppose
that the least power terms in the leading order corrections may be of a nth
power of $\tilde\eta$, then we can eliminate it by a suitable transformation, 
i.e., by completing square, which simply produces a (n-2)th power term in the 
exponent when the corresponding Jacobian factor is exponentiated. This shows 
that
the leading order terms can be transformed away by finite steps to give the
r.h.s. of (20), and completes our systematic approach.

For the sake of practical validity of the method, we applied, in use of suitable computer softwares,
the above algorithm to cases in the flat space and in
the curved space within some orders ${\cal O}(\Delta t^n)$, respectively. 
The results are new as far as we know.

\subsection{Flat space local order 3 algorithm}

In flat spaces i.e. $g_{\mu\nu}=\delta_{\mu\nu}$
the Langevin step algorithms that reproduce the transition density of the
Fokker-Planck equations up to order $\Delta t^3$ are obtained in the Taylor
scheme and in the Runge-Kutta-like scheme, respectively. In the
latter the higher order derivatives are absent.

Our final result in the Taylor scheme is
\begin{eqnarray}
&&\Delta x^i(s,\eta,x_0)=s \eta^i+s^2 u^i+{1\over 2}s^3 \eta^j\partial_j
u^i\nonumber\\&&+s^4 ({1\over 6}\eta^j\eta^k\partial_j\partial_k u^i+{1\over 2}u^j
 \partial_j u^i+{1\over 6}\partial^2 u^i)\nonumber\\
&&+s^5 ({1\over 3}\eta^ju^k \partial_i\partial_ku^i+{1\over 6}\eta^j\partial_j
u^k \partial_k u^i\nonumber\\
&&+{1\over 24}\eta^j\partial_k u^j\partial_ku^i+{1\over 3}\eta^j\partial_j 
\partial^2 u^i)\nonumber \\
&&+s^6({1\over 3}\partial_ju^k\partial_j\partial_ku^i+{1\over 6}u^j u^k
\partial_j\partial_k u^i+{1\over 6}\partial^2u^j \partial_ju^i\nonumber\\
&&+{1\over 6}u^j\partial_ju^k\partial_ku^i+{1\over 3}u^j\partial_j\partial^2
u^i+{1\over 6}\partial^2\partial^2u^i)
\end{eqnarray}  
Here $s=\sqrt{\Delta t}$ and $\eta^i$'s are the Gaussian random variates which
satisfy
\begin{equation}
<\eta^i, \eta^j>=2\delta^{ij}
\end{equation}
Remark that there are terms non-linear in $\eta$'s.

We also derived the $O(\Delta t^3)$ Runge-Kutta-like 
\footnote{ Transcription of (22) to the pure Runge-Kutta scheme is
impossible due to contraction structure of $\eta^j$ and $u^j$ in (22).}
scheme which
contains two sets of the Gaussian random variates $\eta^i$ and $\eta'^j$.

\subsection{ Curved space local order 2 algorithm }

In general positive metric curved spaces, the Langevin step algorithm that
reproduces the transition density of the Fokker-Planck equation up to order
$\Delta t^2$ is obtained in the Taylor scheme.

Our final result is
\begin{eqnarray}
&&\Delta x^i(s,\eta,x_0)=s \theta^i\nonumber\\
&&+s^2 (u^i+{1\over 4}\theta_j\theta^k
 \partial_k g^{ij} -{1\over 2}\partial_j g^{ij})\nonumber \\
&&+s^3 ({1\over 2}\theta^j\partial_j u^i
       + {1\over 4}\theta_k u^j\partial_j g^{ik}
       + {1\over 4}\theta_k g^{mn} \partial_m\partial_n g^{ik}\nonumber \\
&&-{1\over 16}\theta_jg^{km}g_{nl}\partial_kg^{in}\partial_m g^{jl}-{1\over 16}\theta_j\partial_kg^{im}\partial_m g^{jk})\nonumber \\
&&+s^4 ({1\over 2}u^j\partial_ju^i+{1\over 2}g^{jk}\partial_j\partial_ku^i)
\end{eqnarray}  
Here all $g^{ij}$'s, $u^i$'s and their higher derivatives are evaluated at
$x_0$, and $\theta^i$'s are given by

\begin{equation}
\theta^i=\sqrt{g^{ij}(x_0)}\eta_j
\end{equation}
and $\eta_j$'s are the Gaussian random variates as in (23).

As an example, we transcribe the above algorithm to the standard
SU(2) algorithm on ${\bf S}^3$ manifold.
The ${\bf S}^3$ manifold can be patched by two hemispheres and
points ${\bf x}$ on the northern hemisphere are projected stereographically
at ${\bf y}$ on the tangential plane at the north pole and points on the
southern hemisphere are projected on the tangential plane at the south pole.

The $x_4$-component of a point on the northern hemisphere is parametrized
as $x_4=cos\theta$ and ${\bf y}= 2 tan{\theta\over 2}{{\bf x}\over \vert {\bf x}\vert}$. Similar parametrization is done for the southern hemishere.

Using 3 dimensional Gaussian random variates $\eta^i$ of variance 1,
the $\Delta y^i$ can be written as
\begin{eqnarray}
&&\Delta y^i(s,\eta ,y_0)=s \sqrt 2f\eta^i\nonumber\\
&&+s^2[-(4\beta+f)y^i+\eta^i f(y^k \eta^k)/2]\nonumber\\
&&+s^3[\eta^i (-12\beta f+8\beta +{1\over 2}(f^2+f))\nonumber\\&&
-{3\over 8}f(y^k \eta^k)y^i]/\sqrt 2\nonumber\\
&&+s^4{y^i\over 2}[16\beta^2+4(2 f-1)\beta-{1\over 2}(f^2+f)]    
\end{eqnarray}
where $f=1+{\bf y}^2/4$.

The expectation value of $x_4$ is
\begin{equation}
<x_4>={1\over N} \sum_{i=1}^N {{\bf y}^2-4\over {\bf y}^2+4} .
\end{equation}

\section{Numerical examples}

We consider the standard probability distribution
\begin{equation}
p(x)\propto exp(\beta S(x))
\end{equation}
where $x=\theta$ and $S=cos\theta$ in U(1) case, and $S=4 x_4$ with
a four-vector $x_i$ satisfying $x_i^2=1$ in SU(2) case, respectively.

\subsection{ U(1) on ${\bf S}^1$  }

The numerical simulation of the expectation values $<sin^2\theta>$ is
performed, and its data are shown in Fig.1.
The data are taken from successive 2000 Langevin steps
each of which are the average of 100,000 runs.

\begin{figure}[hbt]
\setlength{\unitlength}{0.240900pt}
\ifx\plotpoint\undefined\newsavebox{\plotpoint}\fi
\sbox{\plotpoint}{\rule[-0.200pt]{0.400pt}{0.400pt}}%
\begin{picture}(900,600)(0,0)
\font\gnuplot=cmr10 at 10pt
\gnuplot
\sbox{\plotpoint}{\rule[-0.200pt]{0.400pt}{0.400pt}}%
\put(176.0,113.0){\rule[-0.200pt]{0.400pt}{100.937pt}}
\put(176.0,155.0){\rule[-0.200pt]{4.818pt}{0.400pt}}
\put(154,155){\makebox(0,0)[r]{0.1}}
\put(816.0,155.0){\rule[-0.200pt]{4.818pt}{0.400pt}}
\put(176.0,260.0){\rule[-0.200pt]{4.818pt}{0.400pt}}
\put(154,260){\makebox(0,0)[r]{0.15}}
\put(816.0,260.0){\rule[-0.200pt]{4.818pt}{0.400pt}}
\put(176.0,364.0){\rule[-0.200pt]{4.818pt}{0.400pt}}
\put(154,364){\makebox(0,0)[r]{0.2}}
\put(816.0,364.0){\rule[-0.200pt]{4.818pt}{0.400pt}}
\put(176.0,469.0){\rule[-0.200pt]{4.818pt}{0.400pt}}
\put(154,469){\makebox(0,0)[r]{0.25}}
\put(816.0,469.0){\rule[-0.200pt]{4.818pt}{0.400pt}}
\put(176.0,113.0){\rule[-0.200pt]{0.400pt}{4.818pt}}
\put(176,68){\makebox(0,0){0}}
\put(176.0,512.0){\rule[-0.200pt]{0.400pt}{4.818pt}}
\put(270.0,113.0){\rule[-0.200pt]{0.400pt}{4.818pt}}
\put(270,68){\makebox(0,0){0.05}}
\put(270.0,512.0){\rule[-0.200pt]{0.400pt}{4.818pt}}
\put(364.0,113.0){\rule[-0.200pt]{0.400pt}{4.818pt}}
\put(364,68){\makebox(0,0){0.1}}
\put(364.0,512.0){\rule[-0.200pt]{0.400pt}{4.818pt}}
\put(458.0,113.0){\rule[-0.200pt]{0.400pt}{4.818pt}}
\put(458,68){\makebox(0,0){0.15}}
\put(458.0,512.0){\rule[-0.200pt]{0.400pt}{4.818pt}}
\put(552.0,113.0){\rule[-0.200pt]{0.400pt}{4.818pt}}
\put(552,68){\makebox(0,0){0.2}}
\put(552.0,512.0){\rule[-0.200pt]{0.400pt}{4.818pt}}
\put(646.0,113.0){\rule[-0.200pt]{0.400pt}{4.818pt}}
\put(646,68){\makebox(0,0){0.25}}
\put(646.0,512.0){\rule[-0.200pt]{0.400pt}{4.818pt}}
\put(740.0,113.0){\rule[-0.200pt]{0.400pt}{4.818pt}}
\put(740,68){\makebox(0,0){0.3}}
\put(740.0,512.0){\rule[-0.200pt]{0.400pt}{4.818pt}}
\put(834.0,113.0){\rule[-0.200pt]{0.400pt}{4.818pt}}
\put(834,68){\makebox(0,0){0.35}}
\put(834.0,512.0){\rule[-0.200pt]{0.400pt}{4.818pt}}
\put(176.0,113.0){\rule[-0.200pt]{158.994pt}{0.400pt}}
\put(836.0,113.0){\rule[-0.200pt]{0.400pt}{100.937pt}}
\put(176.0,532.0){\rule[-0.200pt]{158.994pt}{0.400pt}}
\put(506,23){\makebox(0,0){$\Delta t$}}
\put(506,577){\makebox(0,0){$sin^2(\theta)$ Taylor}}
\put(176.0,113.0){\rule[-0.200pt]{0.400pt}{100.937pt}}
\put(706,467){\makebox(0,0)[r]{2nd order}}
\put(750,467){\circle{18}}
\put(185,319){\circle{18}}
\put(195,320){\circle{18}}
\put(214,319){\circle{18}}
\put(232,318){\circle{18}}
\put(251,317){\circle{18}}
\put(270,315){\circle{18}}
\put(364,300){\circle{18}}
\put(458,273){\circle{18}}
\put(552,232){\circle{18}}
\put(646,181){\circle{18}}
\put(740,132){\circle{18}}
\put(834,135){\circle{18}}
\put(728.0,467.0){\rule[-0.200pt]{15.899pt}{0.400pt}}
\put(728.0,457.0){\rule[-0.200pt]{0.400pt}{4.818pt}}
\put(794.0,457.0){\rule[-0.200pt]{0.400pt}{4.818pt}}
\put(185.0,318.0){\rule[-0.200pt]{0.400pt}{0.723pt}}
\put(175.0,318.0){\rule[-0.200pt]{4.818pt}{0.400pt}}
\put(175.0,321.0){\rule[-0.200pt]{4.818pt}{0.400pt}}
\put(195.0,318.0){\rule[-0.200pt]{0.400pt}{0.723pt}}
\put(185.0,318.0){\rule[-0.200pt]{4.818pt}{0.400pt}}
\put(185.0,321.0){\rule[-0.200pt]{4.818pt}{0.400pt}}
\put(214.0,318.0){\rule[-0.200pt]{0.400pt}{0.482pt}}
\put(204.0,318.0){\rule[-0.200pt]{4.818pt}{0.400pt}}
\put(204.0,320.0){\rule[-0.200pt]{4.818pt}{0.400pt}}
\put(232.0,317.0){\rule[-0.200pt]{0.400pt}{0.723pt}}
\put(222.0,317.0){\rule[-0.200pt]{4.818pt}{0.400pt}}
\put(222.0,320.0){\rule[-0.200pt]{4.818pt}{0.400pt}}
\put(251.0,316.0){\rule[-0.200pt]{0.400pt}{0.482pt}}
\put(241.0,316.0){\rule[-0.200pt]{4.818pt}{0.400pt}}
\put(241.0,318.0){\rule[-0.200pt]{4.818pt}{0.400pt}}
\put(270.0,314.0){\rule[-0.200pt]{0.400pt}{0.723pt}}
\put(260.0,314.0){\rule[-0.200pt]{4.818pt}{0.400pt}}
\put(260.0,317.0){\rule[-0.200pt]{4.818pt}{0.400pt}}
\put(364.0,299.0){\rule[-0.200pt]{0.400pt}{0.482pt}}
\put(354.0,299.0){\rule[-0.200pt]{4.818pt}{0.400pt}}
\put(354.0,301.0){\rule[-0.200pt]{4.818pt}{0.400pt}}
\put(458.0,271.0){\rule[-0.200pt]{0.400pt}{0.723pt}}
\put(448.0,271.0){\rule[-0.200pt]{4.818pt}{0.400pt}}
\put(448.0,274.0){\rule[-0.200pt]{4.818pt}{0.400pt}}
\put(552.0,231.0){\rule[-0.200pt]{0.400pt}{0.482pt}}
\put(542.0,231.0){\rule[-0.200pt]{4.818pt}{0.400pt}}
\put(542.0,233.0){\rule[-0.200pt]{4.818pt}{0.400pt}}
\put(646.0,180.0){\rule[-0.200pt]{0.400pt}{0.482pt}}
\put(636.0,180.0){\rule[-0.200pt]{4.818pt}{0.400pt}}
\put(636.0,182.0){\rule[-0.200pt]{4.818pt}{0.400pt}}
\put(740.0,131.0){\rule[-0.200pt]{0.400pt}{0.482pt}}
\put(730.0,131.0){\rule[-0.200pt]{4.818pt}{0.400pt}}
\put(730.0,133.0){\rule[-0.200pt]{4.818pt}{0.400pt}}
\put(834.0,134.0){\rule[-0.200pt]{0.400pt}{0.482pt}}
\put(824.0,134.0){\rule[-0.200pt]{4.818pt}{0.400pt}}
\put(824.0,136.0){\rule[-0.200pt]{4.818pt}{0.400pt}}
\sbox{\plotpoint}{\rule[-0.400pt]{0.800pt}{0.800pt}}%
\put(706,422){\makebox(0,0)[r]{3rd order}}
\put(750,422){\circle*{18}}
\put(185,320){\circle*{18}}
\put(195,320){\circle*{18}}
\put(214,320){\circle*{18}}
\put(232,320){\circle*{18}}
\put(251,320){\circle*{18}}
\put(270,320){\circle*{18}}
\put(364,323){\circle*{18}}
\put(458,331){\circle*{18}}
\put(552,350){\circle*{18}}
\put(646,386){\circle*{18}}
\put(740,448){\circle*{18}}
\put(728.0,422.0){\rule[-0.400pt]{15.899pt}{0.800pt}}
\put(728.0,412.0){\rule[-0.400pt]{0.800pt}{4.818pt}}
\put(794.0,412.0){\rule[-0.400pt]{0.800pt}{4.818pt}}
\put(185.0,318.0){\usebox{\plotpoint}}
\put(175.0,318.0){\rule[-0.400pt]{4.818pt}{0.800pt}}
\put(175.0,321.0){\rule[-0.400pt]{4.818pt}{0.800pt}}
\put(195.0,318.0){\usebox{\plotpoint}}
\put(185.0,318.0){\rule[-0.400pt]{4.818pt}{0.800pt}}
\put(185.0,321.0){\rule[-0.400pt]{4.818pt}{0.800pt}}
\put(214.0,318.0){\usebox{\plotpoint}}
\put(204.0,318.0){\rule[-0.400pt]{4.818pt}{0.800pt}}
\put(204.0,321.0){\rule[-0.400pt]{4.818pt}{0.800pt}}
\put(232.0,318.0){\usebox{\plotpoint}}
\put(222.0,318.0){\rule[-0.400pt]{4.818pt}{0.800pt}}
\put(222.0,321.0){\rule[-0.400pt]{4.818pt}{0.800pt}}
\put(251.0,318.0){\usebox{\plotpoint}}
\put(241.0,318.0){\rule[-0.400pt]{4.818pt}{0.800pt}}
\put(241.0,321.0){\rule[-0.400pt]{4.818pt}{0.800pt}}
\put(270.0,319.0){\usebox{\plotpoint}}
\put(260.0,319.0){\rule[-0.400pt]{4.818pt}{0.800pt}}
\put(260.0,322.0){\rule[-0.400pt]{4.818pt}{0.800pt}}
\put(364.0,321.0){\usebox{\plotpoint}}
\put(354.0,321.0){\rule[-0.400pt]{4.818pt}{0.800pt}}
\put(354.0,324.0){\rule[-0.400pt]{4.818pt}{0.800pt}}
\put(458.0,330.0){\usebox{\plotpoint}}
\put(448.0,330.0){\rule[-0.400pt]{4.818pt}{0.800pt}}
\put(448.0,333.0){\rule[-0.400pt]{4.818pt}{0.800pt}}
\put(552.0,349.0){\usebox{\plotpoint}}
\put(542.0,349.0){\rule[-0.400pt]{4.818pt}{0.800pt}}
\put(542.0,352.0){\rule[-0.400pt]{4.818pt}{0.800pt}}
\put(646.0,385.0){\usebox{\plotpoint}}
\put(636.0,385.0){\rule[-0.400pt]{4.818pt}{0.800pt}}
\put(636.0,388.0){\rule[-0.400pt]{4.818pt}{0.800pt}}
\put(740.0,446.0){\rule[-0.400pt]{0.800pt}{0.964pt}}
\put(730.0,446.0){\rule[-0.400pt]{4.818pt}{0.800pt}}
\put(730.0,450.0){\rule[-0.400pt]{4.818pt}{0.800pt}}
\end{picture}
\caption  {The results of simulation of $<sin^2\theta>$ in U(1) system. 
$\beta=5$, standard method in the Taylor scheme.}

\end{figure}
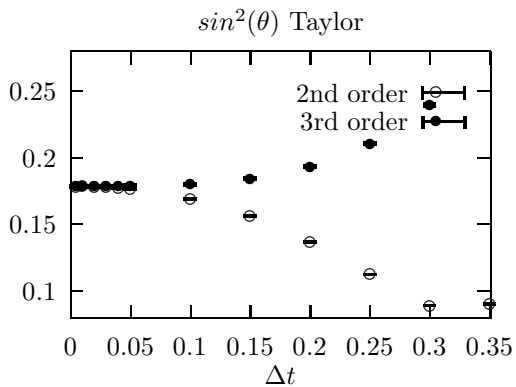

\subsection{ SU(2) on ${\bf S}^3$  }

We performed numerical simulation of the expectation values $<x_4>$ for
$\beta=1$, results of which are shown in Fig.2.
Data are taken from successive Langevin times from $t=2$ to $t=12$ each of 
which are the average of 100,000 runs.  

\begin{figure}[hbt]
\setlength{\unitlength}{0.240900pt}
\ifx\plotpoint\undefined\newsavebox{\plotpoint}\fi
\sbox{\plotpoint}{\rule[-0.200pt]{0.400pt}{0.400pt}}%
\begin{picture}(900,600)(0,0)
\font\gnuplot=cmr10 at 10pt
\gnuplot
\sbox{\plotpoint}{\rule[-0.200pt]{0.400pt}{0.400pt}}%
\put(176.0,113.0){\rule[-0.200pt]{0.400pt}{100.937pt}}
\put(176.0,113.0){\rule[-0.200pt]{4.818pt}{0.400pt}}
\put(154,113){\makebox(0,0)[r]{0.6}}
\put(816.0,113.0){\rule[-0.200pt]{4.818pt}{0.400pt}}
\put(176.0,218.0){\rule[-0.200pt]{4.818pt}{0.400pt}}
\put(154,218){\makebox(0,0)[r]{0.65}}
\put(816.0,218.0){\rule[-0.200pt]{4.818pt}{0.400pt}}
\put(176.0,323.0){\rule[-0.200pt]{4.818pt}{0.400pt}}
\put(154,323){\makebox(0,0)[r]{0.7}}
\put(816.0,323.0){\rule[-0.200pt]{4.818pt}{0.400pt}}
\put(176.0,427.0){\rule[-0.200pt]{4.818pt}{0.400pt}}
\put(154,427){\makebox(0,0)[r]{0.75}}
\put(816.0,427.0){\rule[-0.200pt]{4.818pt}{0.400pt}}
\put(176.0,532.0){\rule[-0.200pt]{4.818pt}{0.400pt}}
\put(154,532){\makebox(0,0)[r]{0.8}}
\put(816.0,532.0){\rule[-0.200pt]{4.818pt}{0.400pt}}
\put(176.0,113.0){\rule[-0.200pt]{0.400pt}{4.818pt}}
\put(176,68){\makebox(0,0){0}}
\put(176.0,512.0){\rule[-0.200pt]{0.400pt}{4.818pt}}
\put(333.0,113.0){\rule[-0.200pt]{0.400pt}{4.818pt}}
\put(333,68){\makebox(0,0){0.05}}
\put(333.0,512.0){\rule[-0.200pt]{0.400pt}{4.818pt}}
\put(490.0,113.0){\rule[-0.200pt]{0.400pt}{4.818pt}}
\put(490,68){\makebox(0,0){0.1}}
\put(490.0,512.0){\rule[-0.200pt]{0.400pt}{4.818pt}}
\put(647.0,113.0){\rule[-0.200pt]{0.400pt}{4.818pt}}
\put(647,68){\makebox(0,0){0.15}}
\put(647.0,512.0){\rule[-0.200pt]{0.400pt}{4.818pt}}
\put(805.0,113.0){\rule[-0.200pt]{0.400pt}{4.818pt}}
\put(805,68){\makebox(0,0){0.2}}
\put(805.0,512.0){\rule[-0.200pt]{0.400pt}{4.818pt}}
\put(176.0,113.0){\rule[-0.200pt]{158.994pt}{0.400pt}}
\put(836.0,113.0){\rule[-0.200pt]{0.400pt}{100.937pt}}
\put(176.0,532.0){\rule[-0.200pt]{158.994pt}{0.400pt}}
\put(506,23){\makebox(0,0){$\Delta t$}}
\put(506,577){\makebox(0,0){$SU2\quad  S^3\quad  \beta=1$}}
\put(176.0,113.0){\rule[-0.200pt]{0.400pt}{100.937pt}}
\put(706,467){\makebox(0,0)[r]{$<x_4>$}}
\put(750,467){\circle{18}}
\put(176,235){\circle{18}}
\put(192,235){\circle{18}}
\put(207,235){\circle{18}}
\put(333,244){\circle{18}}
\put(490,270){\circle{18}}
\put(805,358){\circle{18}}
\put(728.0,467.0){\rule[-0.200pt]{15.899pt}{0.400pt}}
\put(728.0,457.0){\rule[-0.200pt]{0.400pt}{4.818pt}}
\put(794.0,457.0){\rule[-0.200pt]{0.400pt}{4.818pt}}
\put(176,235){\usebox{\plotpoint}}
\put(166.0,235.0){\rule[-0.200pt]{4.818pt}{0.400pt}}
\put(166.0,235.0){\rule[-0.200pt]{4.818pt}{0.400pt}}
\put(192.0,233.0){\rule[-0.200pt]{0.400pt}{0.723pt}}
\put(182.0,233.0){\rule[-0.200pt]{4.818pt}{0.400pt}}
\put(182.0,236.0){\rule[-0.200pt]{4.818pt}{0.400pt}}
\put(207.0,233.0){\rule[-0.200pt]{0.400pt}{0.964pt}}
\put(197.0,233.0){\rule[-0.200pt]{4.818pt}{0.400pt}}
\put(197.0,237.0){\rule[-0.200pt]{4.818pt}{0.400pt}}
\put(333.0,242.0){\rule[-0.200pt]{0.400pt}{0.723pt}}
\put(323.0,242.0){\rule[-0.200pt]{4.818pt}{0.400pt}}
\put(323.0,245.0){\rule[-0.200pt]{4.818pt}{0.400pt}}
\put(490.0,268.0){\rule[-0.200pt]{0.400pt}{0.723pt}}
\put(480.0,268.0){\rule[-0.200pt]{4.818pt}{0.400pt}}
\put(480.0,271.0){\rule[-0.200pt]{4.818pt}{0.400pt}}
\put(805.0,356.0){\rule[-0.200pt]{0.400pt}{0.723pt}}
\put(795.0,356.0){\rule[-0.200pt]{4.818pt}{0.400pt}}
\put(795.0,359.0){\rule[-0.200pt]{4.818pt}{0.400pt}}
\end{picture}
\caption  {The results of simulation of $<x_4>$  of SU(2) in
${\bf S}^3$.  $\beta=1$  }
\end{figure}
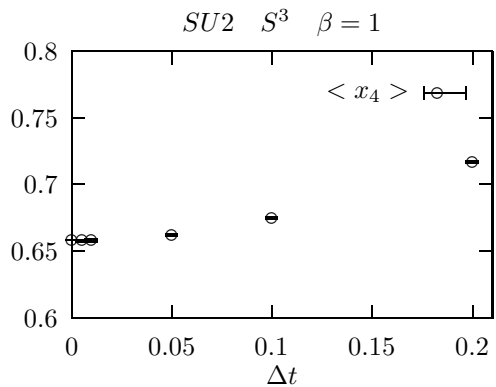

\section{Conclusion and Outlook} 

General algorithm for obtaining higher order corrections to the Langevin
step algorithms with respect to $\Delta t$ was formulated. 
The method is straightforward without matching equations
and is tractable by computer.

A local third order algorithm for simulating general
flat space Langevin equations and a local second order algorithm for simulating general curved space Langevin equations were
obtained.

This method is based on Gaussian distribution in use of noncompact
coordinates, as its lowest leading order approximation to transition density.
Even in case of compact curved spaces as ${\bf S}^n$, however, patching
algorithm in use of stereographic projection on two noncompact tangential
planes turned out to be successful and fast on computer.
Similar method is applicable in principle to simulation of
imaginary time evolution of wave functions in quantum mechanics.

\end{document}